\documentclass{elsart}

\usepackage{graphicx} 
\usepackage{epsfig} 

\usepackage{subfigure}

\usepackage{amssymb}

\usepackage{bm}

\begin{document}

\begin{frontmatter}



\title{\boldmath
Observation of charged $\kappa$ in 
$J/\psi\rightarrow K^{*}(892)^{\mp}K_{S}\pi^{\pm},\ K^{*}(892)^{\mp}\rightarrow K_{S}\pi^{\mp}$ at BESII
}
\date{}

\vspace{-1.5em}

{\normalsize 25 August, 2010; Revised 2 February, 2011}

\begin{small}
\begin{center}
M.~Ablikim$^{1}$,              J.~Z.~Bai$^{1}$,   Y.~Bai$^{1}$,
Y.~Ban$^{15}$,
X.~Cai$^{1}$,                  H.~F.~Chen$^{22}$,
H.~S.~Chen$^{1}$,              J.~C.~Chen$^{1}$,
Jin~Chen$^{1}$,
Y.~B.~Chen$^{1}$
% ...
\\(BES Collaboration)
{\it

$^{1}$ Institute of High Energy Physics, Beijing 100049, People's Republic of China\\
% ...
$^{15}$ Peking University, Beijing 100871, People's Republic of China\\
% ...
$^{22}$ University of Science and Technology of China, Hefei 230026,
People's Republic of China\\
% ...
}
\end{center}
\end{small}

\normalsize
%
\begin{abstract}
  
  Using $58$ million $ J/\psi$ decays obtained by BESII, a charged
  $\kappa$ particle is observed in the analysis of the
  $K_{S}\pi^{\pm}$ system recoiling against $K^{*}(892)^{\mp}$
  selected in $J/\psi \to K_{S}K_{S}\pi^{+}\pi^{-}$.  The mass and
  width values of the charged $\kappa$ are obtained to be $(826\pm
  49_{-34}^{+49})$ MeV/$c^{2}$ and $(449\pm 156_{-81}^{+144})$
  MeV/$c^{2}$ for the Breit-Wigner parameters, and the pole position
  is determined to be $(764\pm
  63_{-54}^{+71})-i(306\pm 149_{-85}^{+143})$ MeV/$c^{2}$.  They are
  in good agreement with those of the neutral $\kappa$ observed by the
  BES Collaboration.
\end{abstract}

\begin{keyword}
Charged $\kappa$, low mass scalar, $J/ \psi$ decays, $K^*(892) K \pi$
%
\end{keyword}
\end{frontmatter}



The experimental evidence on the existence of the low mass scalars
below 1 GeV/$c^{2}$, $f_{0}(600)/\sigma$ and $ K_{0}^{*}(800)/\kappa$
particles \cite{1}, have stimulated studies on the lower mass meson
structures.  They may be classified in a possible $\sigma$ nonet.
However their nature is still controversial.  The neutral $\kappa$
particle was observed \cite{2,3,4,5,6,7,8,9,10,11,12a} in the analyses of
$K\pi$ scattering data.
It has also been confirmed in the production processes, $D^{+}$ decay
into $K^{-}\pi^{+}\pi^{+}$ by E791 at Fermilab \cite{13}, and $
J/\psi$ decay into $\overline{K}^{*}(892)^{0}K^{+}\pi^{-}$ by BESII
\cite{12}.  The existence of a coherent $ K\pi$ $S$-wave contribution
was presented by FOCUS \cite{14} in the analysis of semileptonic $D$
decay, $D^{+}\rightarrow K^{-}\pi^{+}\mu^{+}\nu$.  Recently the
necessity of the $\kappa$ was confirmed in the Dalitz plot analysis of
the hadronic $D$ decay, $D^{+}\rightarrow K^{-}\pi^{+}\pi^{+} $ also
by FOCUS \cite{15} and later in the alternate analysis on the same
data \cite{16}.  The CLEO Collaboration also confirmed the neutral
$\kappa$ in the same channel \cite{17}.  We may expect the existence
of a charged $\kappa$ according with isospin symmetry.


As for the charged $\kappa$, CLEO reported \cite{18} the necessity of
a $S$-wave $K^{\pm}\pi^{0}$ resonance in the analysis of the
interfering $K^{*}(892)^+K^{-}$ and $K^{*}(892)^-K^{+}$ amplitudes in
the decay, $D^{0}\rightarrow K^{+}K^{-}\pi^{0}$.  Recently Belle
\cite{19} and BaBar \cite{20} reported the necessity of the charged
$\kappa$ in the analyses of the $K_{S}\pi^{-}$ mass distribution in
the semileptonic $\tau$ decay, $\tau^{-}\rightarrow
K_{S}\pi^{-}\nu_{\tau}$.  BaBar reported, however, no need for the
charged $\kappa$ in their analysis of the $D$ decay into
$K^{+}K^{-}\pi^{0}$ \cite{21}.  BES reported recently
the charged $\kappa$ in the partial wave analysis (PWA) of the
combined system, $K^{*}(892)^{\mp}K^{\pm}\pi^{0}$ and
$K^{*}(892)^{\mp}K_{S}\pi^{\pm}$, in the decay, $J/\psi\rightarrow
K^{\pm}K_{S}\pi^{\mp}\pi^{0}$ \cite{22}.  The results for its
resonance parameters are consistent with those of the neutral $\kappa$
\cite{12}.


In this report, we present the results of a PWA of the
$K^{*}(892)^{\mp}K_{S}\pi^{\pm}$ system in the decay
$J/\psi\rightarrow K^{*}(892)^{\mp}K_{S}\pi^{\pm},
K^{*}(892)^{\mp}\rightarrow K_{S}\pi^{\mp}$, based on 58 million $
J/\psi$ decays collected by BESII at BEPC (Beijing Electron Positron
Collider).  The BESII detector is described in detail elsewhere
\cite{23}.  
The resonance parameters of the charged $\kappa$ are obtained. 
In the PWA, a Breit-Wigner parameterization with an
s-dependent width is used, and parameters for its pole position are
determined.  A different parameterization with a constant width is
also examined.  A branching ratio for $J/\psi\rightarrow
K^{*}(892)^{\mp}\kappa^{\pm}$ is determined.


In the event selection, the following requirements are imposed. The
event must have six charged tracks with zero net charge.  Each charged
track must have a good helix fit in the main drift chamber (MDC) and
satisfy for the polar angle $\theta$, $\left|\cos\theta\right|<0.8$
and for the transverse momentum $P_{t},\ P_{t}>50 \ 
\mathrm{MeV}/c^{2}$ in the MDC.  Two $K_{S} \to \pi^+ \pi^-$
candidates, $\left(\pi^{+}\pi^{-}\right)^{(1)}$ or
$\left(\pi^{+}\pi^{-}\right)^{(2)}$, having a minimum value for
 $\delta m_{K_{S}}$,
\begin{eqnarray}
  \delta m_{K_{S}} \equiv \sqrt{
    (m_{\pi^{+}\pi^{-}}^{(1)} - m_{K_{S}})^{2} + 
    (m_{\pi^{+}\pi^{-}}^{(2)}-m_{K_{S}})^{2}} ,
  \label{eq1}
\end{eqnarray}
%
\begin{figure}[thbp]
\begin{center}
\includegraphics[width=10cm,angle=0]{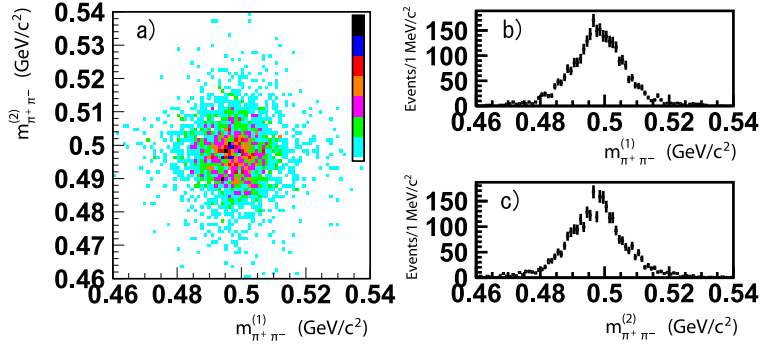}
\end{center}
\caption{
   {\footnotesize 
     Scatter plot for (a) $m_{\pi^{+}\pi^{-}}^{(1)}$ versus
     $m_{\pi^{+}\pi^{-}}^{(2)}$ and its projected spectra for (b)
     $m_{\pi^{+}\pi^{-}}^{(1)}$ and (c) $m_{\pi^{+}\pi^{-}}^{(2)}$. 
     The different colors on the color scale in (a) correspond to 
     increments of two in the number of events in the bin. 
   }  
}
\label{fig:1}
\end{figure}
where $m_{\pi^{+}\pi^{-}}^{(i)}\ (i=1$ or $2)$ is the invariant mass
of the $(\pi^{+}\pi^{-})^{(i)}$ pair, are selected, and 
$\delta m_{K_{S}}$ is required to satisfy $\delta m_{K_{S}}\leq 0.02$
GeV/$c^{2}$.  The surviving events are fitted kinematically under the
hypothesis, $J/\psi \rightarrow 3(\pi^{+}\pi^{-})$ and are required to have $\chi^{2}$
value less than $15,\ \chi_{6\pi}^{2}<15$.  The positions of closest
approach to the beam axis of the two pions of each
$\left(\pi^{+}\pi^{-}\right)^{(i)}$ pair must agree within $0.05 \ \mbox{m}$ 
along the z axis (the electron beam direction).  

The scatter plot of $m_{\pi^{+}\pi^{-}}^{(1)}$ versus
$m_{\pi^{+}\pi^{-}}^{(2)}$ and its projected spectra for
$m_{\pi^{+}\pi^{-}}^{(1)}$ and $m_{\pi^{+}\pi^{-}}^{(2)}$ are shown in
Fig. \ref{fig:1}.  

The selected $m_{\pi^{+}\pi^{-}}^{(i)}$ event $(i=1 \ \mathrm{or}\ 2)$ is assigned to
be $K_{S}^{(i)}$ in the $K_{S}K_{S}\pi^{+}\pi^{-}$ events.  After the
selection, $2,933$ $K_{S}K_{S}\pi^{+}\pi^{-}$ events survive.

\begin{figure}[htbp]
\begin{center}
\epsfxsize=9cm \epsffile{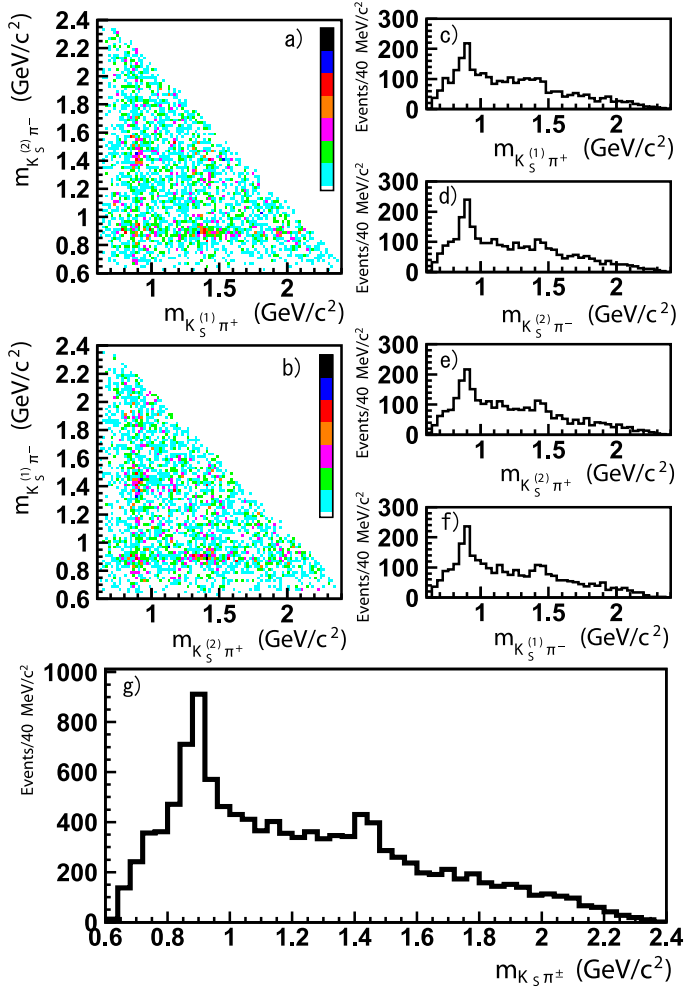}
\end{center}
\vspace{-0.5em}
\caption{
   {\footnotesize 
     Scatter plots for (a) $m_{K_{S}^{(1)}\pi^{+}}$ versus
     $m_{K_{S}^{(2)}\pi^{-}}$ and (b) $m_{K_{S}^{(2)}\pi^{+}}$ versus
     $m_{K_{S}^{(1)}\pi^{-}}$ and their projected spectra for (c)
     $m_{K_{S}^{(1)}\pi^{+}}$, (d) $m_{K_{S}^{(2)}\pi^{-}}$, (e)
     $m_{K_{S}^{(2)}\pi^{+}}$, (f) $m_{K_{S}^{(1)}\pi^{-}}$ and (g)
     their sum, $m_{K_{S}\pi^{\pm}}$.  
     The different colors on the color scale in (a) and (b) correspond to 
     increment of one in the number of event in the bin. 
}  }
\label{fig:2}
\end{figure}

\begin{figure}[htbp]
\begin{center}
\epsfxsize=8.5cm \epsffile{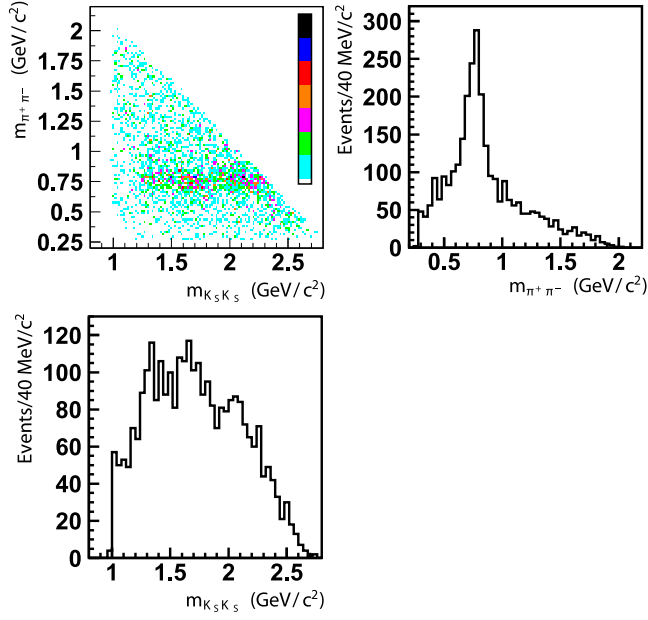}
\end{center}
\vspace{-0.5em}
\caption{
   {\footnotesize 
     Scatter plot for $m_{K_{S}K_{S}}$ versus $m_{\pi^{+}\pi^{-}}$ and
     its projected spectra for $m_{K_{S}K_{S}}$ and
     $m_{\pi^{+}\pi^{-}}$.  
     The different colors on the color scale in the scatter plot correspond to 
     increment of one in the number of event in the bin. 
}  }
\label{fig:3}
\end{figure}

The scatter plots of the surviving events are displayed in
Figs.~\ref{fig:2}a) and \ref{fig:2}b) for the invariant mass of
$K_{S}^{(1)}\pi^{+},\ m_{K_{S}^{(1)}\pi^{+}}$ versus that of
$K_{S}^{(2)}\pi^{-}$, $m_{K_{S}^{(2)}\pi^{-}}$ and for those of the
alternate combinations, $m_{K_{S}^{(2)}\pi^{+}}$ versus
$m_{K_{S}^{(1)}\pi^{-}}$, respectively.  The projected spectra are
shown for $K_{S}^{(1)}\pi^{+}$ (Fig. \ref{fig:2}c)) and
$K_{S}^{(2)}\pi^{-}$ (Fig. \ref{fig:2}d)), and for those of alternate
combinations, $K_{S}^{(2)}\pi^{+}$ (Fig. \ref{fig:2}e)) and
$K_{S}^{(1)}\pi^{-}$ (Fig. \ref{fig:2}f)).  Their sum spectrum is also
shown in Fig. \ref{fig:2}g). Concentrations of events coming from
$K^{*}(892)^{\pm}$ are seen in the spectra.  The scatter plot
for the invariant mass of $K_{S}K_{S}$ versus that of $\pi^{+}\pi^{-}$
is shown with its projected spectra in Fig. \ref{fig:3}.  A
$\rho(770)$ peak is also seen in the $\pi^{+}\pi^{-}$ spectrum.



Before the event selection of $K^{*}(892)^\mp K_{S}\pi^{\pm}$ for the
PWA, other possible backgrounds are studied.  Selection efficiencies
are determined by Monte Carlo simulation for the decays,
$J/\psi\rightarrow 3(\pi^{+}\pi^{-})$, $J/\psi\rightarrow
K^{+}K_{S}\pi^{-}\pi{}^{+}\pi^{-}$, $J/\psi\rightarrow\gamma\eta_{c},\ 
\eta_{c}\rightarrow K_{S}K_{S}\pi^{+}\pi^{-}$,
$J/\psi\rightarrow\gamma K_{S}K_{S}\pi^{+}\pi^{-}$,
$J/\psi\rightarrow\gamma K^{*}(892)^{\mp}K^{*}(892)^{\pm}$,
$J/\psi\rightarrow\gamma K^{*}(892)^\mp K_{S}\pi^{\pm}$ and
$J/\psi\rightarrow\gamma 3(\pi^{+}\pi^{-})$ which could contribute
background to the study of $J/\psi\rightarrow
K^{*}(892)^{\mp}K_{S}\pi^{\pm}\rightarrow K_{S}K_{S}\pi^{+}\pi^{-}$.
Other decay modes such as $J/\psi\rightarrow 3(\pi^{+}\pi^{-})\pi^{0}$
and $J/\psi\rightarrow K_{S}K_{S}\pi^{+}\pi^{-}\pi^{0}$ are also
studied.  The largest background is from $J/\psi\rightarrow
3(\pi^{+}\pi^{-})$.  Those for $J/\psi\rightarrow\gamma\eta_{c},\ 
\eta_{c}\rightarrow K_{S}K_{S}\pi^{+}\pi^{-}$,
$J/\psi\rightarrow\gamma K_{S}K_{S}\pi^{+}\pi^{-}$,
$J/\psi\rightarrow\gamma K^{*}(892)\overline{K}^{*}(892)$ and
$J/\psi\rightarrow\gamma K^{*}(892)^{\pm}K_{S}\pi^{\mp}$ are small.
The last two processes which have $K^{*}(892)$ ($K^{*}(892)$ and
$\overline{K}^{*}(892)$) are suppressed and negligible.  Others are
also found to be negligible.

The contribution of background events from $J/\psi \rightarrow
3(\pi^{+}\pi^{-})$ is checked using both the distributions of $\delta
m_{K_{S}}$ and the proper time$,\ c\tau$ of $K_{S}$.  The $\delta
m_{K_{S}}$ distribution of data shows an excess of events compared
with $K_{S}K_{S}\pi^{+}\pi^{-}$ Monte Carlo simulation outside 
the selection region, $\delta m_{K_{S}}>20$ MeV$/c^{2}$. 
Fitting the excess with parameters obtained
for $3(\pi^{+}\pi^{-})$ by Monte Carlo simulation and extrapolating
the fit down into the selection region gives a contribution from
$3(\pi^{+}\pi^{-})$ to be $(8.7\pm 0.1) \%$.  Next, the $ c\tau$
distribution is studied.  There is an excess of events for $c\tau$
less than 1 cm compared with what is expected from Monte Carlo
simulation of $K_{S}K_{S}\pi^{+}\pi^{-}$ events.  The excess events
are assumed to be $3(\pi^{+}\pi^{-})$, and their amount is estimated
to be $(11.9\pm 0.8) \%$ of the total $K_{S}$ events.  Though the two
values differ, they give an indication of the amount of background
from the $3(\pi^{+}\pi^{-})$ events.  In the PWA, $K^{*}(892)$
side-band events are used for the background estimation, and they may
include not only those from $3(\pi^{+}\pi^{-})$ but also from other
processes, as described below.

Four combinations of $K_{S}^{(i=1\ \mathrm{o}\mathrm{r}\ 2)}\pi^{(+\ 
  \mathrm{o}\mathrm{r}\ -)}$ events recoiling against $K^{*}(892)^{(-\ 
  \mathrm{o}\mathrm{r}\ +)}$ are selected in $J/\psi\rightarrow
K_{S}K_{S}\pi^{+}\pi^{-}$ for the PWA.
Preceding the selection, background events associated with $\rho(770)^{0}$ are rejected with the condition, $\left|m_{\pi^{+}\pi^{-}}-775\right|>100$ MeV/$c^{2}$. 
The conditions
for $K^{*}(892)$ selection and background rejection of
alternate combination channels in the four
combinations are as follows;
\begin{itemize}
\item[(1)] $ J/\psi\rightarrow K^{*}(892)^{-}K_{S}^{(1)}\pi^{+}:\ |m_{K_{S}^{(2)}\pi^{-}}-892|<80$ MeV/$c^{2},\\
 |m_{K_{S}^{(1)}\pi^{-}}-892|>40$ MeV/$c^{2},\ |m_{K_{S}^{(2)}\pi^{+}}-892|>40$ MeV/$c^{2}$, 

\item[(2)] $ J/\psi\rightarrow K^{*}(892)^{+}K_{S}^{(2)}\pi^{-}:\ |m_{K_{S}^{(1)}\pi^{+}}-892|<80$ MeV/$c^{2},\\ 
|m_{K_{S}^{(2)}\pi^{+}}-892|>40$ MeV/$c^{2},\ |m_{K_{S}^{(1)}\pi^{-}}-892|\ >40$ MeV/$c^{2}$,
\item[(3)] $ J/\psi\rightarrow K^{*}(892)^{-}K_{S}^{(2)}\pi^{+}:\ |m_{K_{S}^{(1)}\pi^{-}}-892|<80$ MeV/$c^{2},\\ 
|m_{K_{S}^{(2)}\pi^{-}}-892|>40$ MeV/$c^{2},\ |m_{K_{S}^{(1)}\pi^{+}}-892|>40$ MeV/$c^{2}$, 

\item[or]  

\item[(4)] $ J/\psi\rightarrow K^{*}(892)^{+}K_{S}^{(1)}\pi^{-}:\ |m_{K_{S}^{(2)}\pi^{+}}-892|<80$ MeV/$c^{2},\\ 
|m_{K_{S}^{(1)}\pi^{+}}-892|>40$ MeV/$c^{2},\ |m_{K_{S}^{(2)}\pi^{-}}-892|\ >40$ MeV/$c^{2}$.
\end{itemize}

\begin{figure}[htbp]
\includegraphics[width=14cm,angle=0]{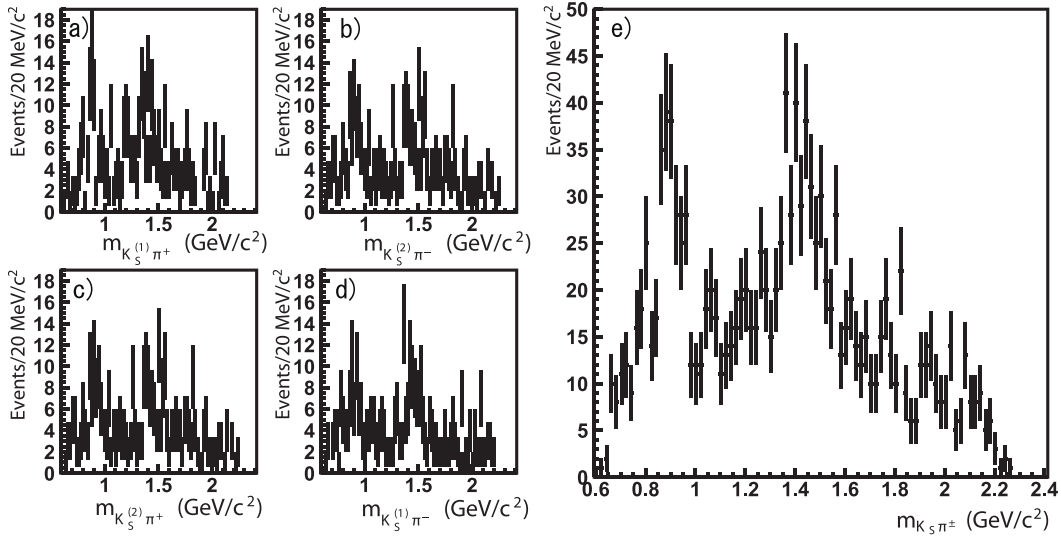}
\vspace{-1.5em}
\caption{
   {\footnotesize 
$m_{K_{S}\pi^{\pm}}$ distributions for (a) $m_{K_{S}^{(1)}\pi^{+}}$, (b) $m_{K_{S}^{(2)}\pi^{-}}$, (c) $m_{K_{S}^{(2)}\pi^{+}}$ and (d) $m_{K_{S}^{(1)}\pi^{-}}$ and (e) their sum, $m_{K_{S}\pi^{\pm}}$.
   }
}
\label{fig:4}
\end{figure}
\begin{figure}[htbp]
\includegraphics[width=14cm,angle=0]{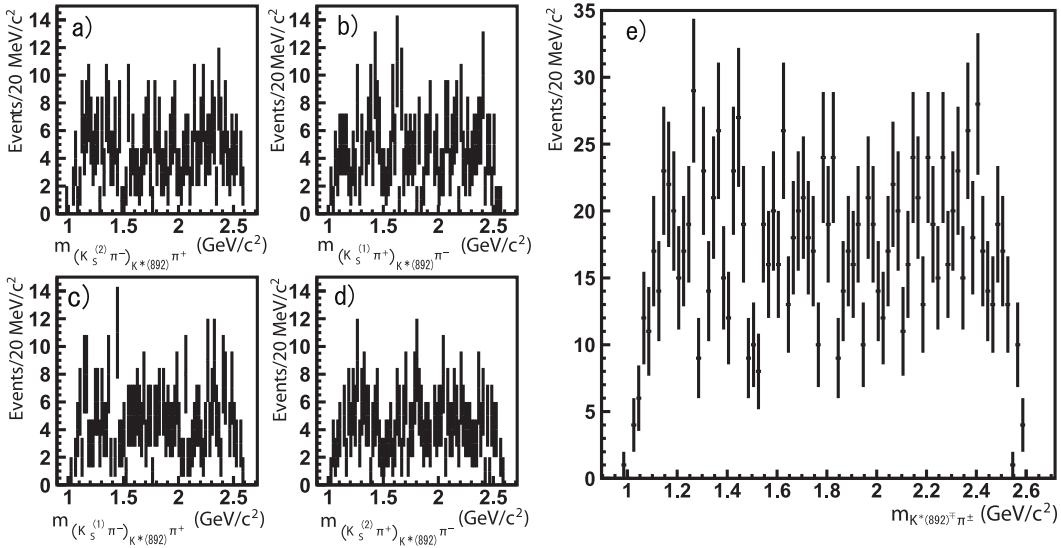}
\vspace{-1.5em}
\caption{
   {\footnotesize 
$m_{K^{*}(892)^{\mp} \pi^{\pm}}$ distributions for 
(a) $m_{(K_{S}^{(2)}\pi^{-})_{K^{*}(892)} \pi^{+}}$, 
(b) $m_{(K_{S}^{(1)}\pi^{+})_{K^{*}(892)} \pi^{-}}$, 
(c) $m_{(K_{S}^{(1)}\pi^{-})_{K^{*}(892)} \pi^{+}} $, 
(d) $m_{(K_{S}^{(2)}\pi^{+})_{K^{*}(892)} \pi^{-}}$ and 
(e) their sum, $m_{K^{*}(892)^{\mp}\pi^{\pm}}$. 
   }
}
\label{fig:5}
\end{figure}
After selection, $1,338$ events survive for the sum of the four
channels and are used in the PWA.
\begin{figure}[htbp]
\begin{center}
\includegraphics[width=5.5cm,angle=0]{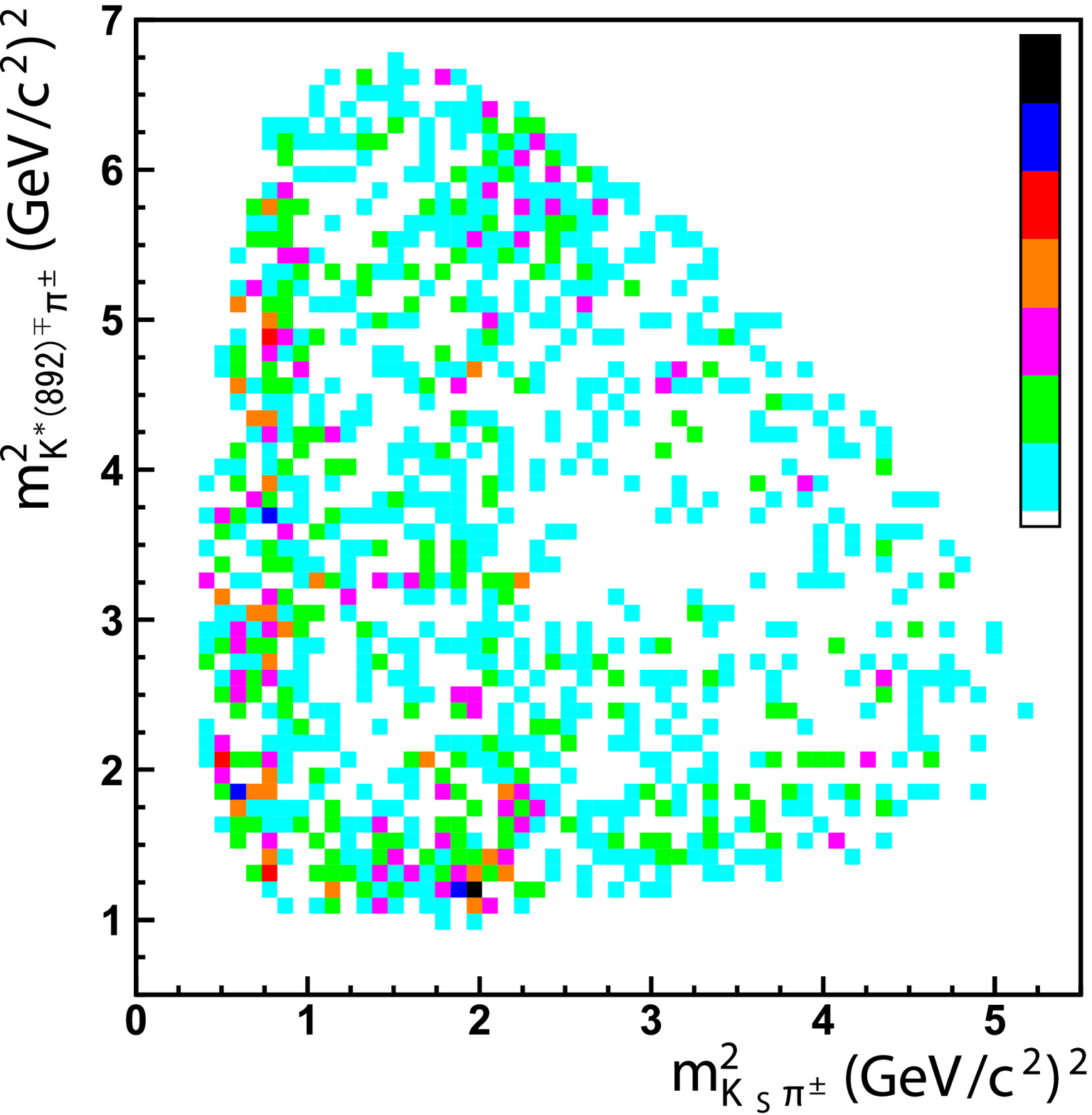}
\end{center}
\caption{
   {\footnotesize 
Dalitz plot of $K_{S}\pi^{\pm}$ versus $K^{*}(892)^{\mp}\pi^{\pm}$. 
The different colors on the color scale correspond to 
increment of one in the number of event in the bin. 
   }
}
\label{fig:6}
\end{figure}
\begin{figure}[htbp]
\begin{center}
\includegraphics[width=5.5cm,angle=0]{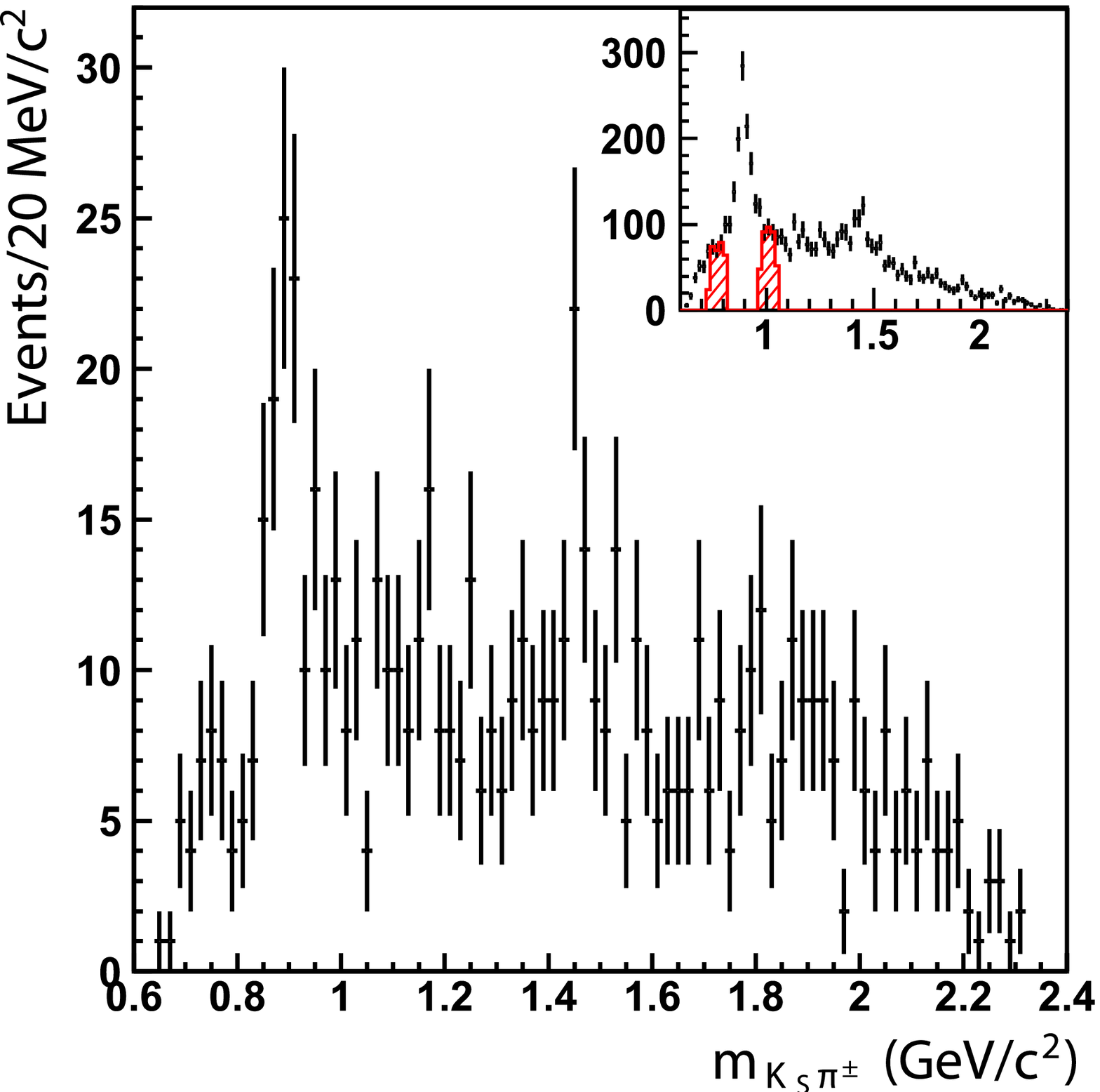}
\end{center}
\caption{
   {\footnotesize 
     $m_{K_{S}\pi^{\pm}}$ distribution of $K^{*}(892)^{\mp}$ side-band
     events. Insertion: $m_{K_{S}\pi}$ distribution after final
     selection except the $K^{*}(892)$ requirement. The hatched areas
     (red color) show the region for the side-band selection.  }  }
\label{fig:7}
\end{figure}

The $m_{K_{S}\pi}$ distributions are shown for the four
$K^{*}(892)^{\mp}K_{S}\pi^{\pm}$ combinations in Figs. \ref{fig:4}a)-
\ref{fig:4}d) and for their sum in Fig. \ref{fig:4}e). Clear peaks
around $900$ MeV/$c^{2}$ and around $1400$ MeV/$c^{2}$ are observed
in the sum distribution. The $m_{K^{*\mp}\pi^{\pm}}$ distributions are
shown in Fig. \ref{fig:5}. 
No distinct structure is seen in the figures. 
$K_{1}(1270)$ which preferentially decays into $\rho K_{S}$ as a background is reduced. 
The Dalitz plot of the summed events is shown 
for $K_{S}\pi^{\pm}$ and $K^{*}(892)^{\mp}\pi^{\pm}$ in Fig. \ref{fig:6}.
Two vertical bands correspond to the peak around 892 MeV/$c^{2}$ and around 1430 MeV/$c^{2}$.

The $m_{K_{S}\pi^{\pm}}$ distribution of the $K^{*}(892)^{\mp}$
side-band events selected in the region, $80$
MeV/$c^{2}<|M_{K^{*}(892)}-892|<160$ MeV/$c^{2}$ is shown in Fig.
\ref{fig:7}. The $m_{K_{S}\pi}$ distribution after the final selection
but without the $K^{*}(892)$ requirement is inserted in the figure.
The side band selection regions are indicated by the hatched area.
The following 
processes, $K^{*}(892)K_{0}^{*}(1430)$, $K^{*}(892)K_{2}^{*}(1430)$,
$K_{0}^{*}(1430)K_{0}^{*}(1430)$, $K_{0}^{*}(1430)\kappa$,
$K_{1}(1270)K_{S}$, $K_{1}(1400)K_{S}$ and $K^{*}(892)K_{S}\pi$ from
alternate combination channels and a background process,
$3(\pi^{+}\pi^{-})$, contribute to the side band events.

The contribution from a non-interfering phase space-like background is estimated 
from the $K^{*}(892)^{\mp}$ side-band events in the $K_{S}\pi^{\pm}$ mass region 
between 1.6 GeV/$c^{2}$ and 2.0 GeV/$c^{2}$ where 
these events are assumed to dominate. 
Uncertainties for the estimation of this contribution will be
considered in the PWA and taken into the estimation for the systematic
errors of the $\kappa$ parameters.


The variant mass and width method (VMW method) \cite{24} is used for
the analysis.  The VMW method is a covariant field-theoretical
approach consistent with generalized unitarity.  In this method, the
total amplitude is expressed as a coherent sum of respective
amplitudes corresponding to the relevant processes of strong
interactions among all color singlet hadrons.  As the bases of
S-matrix for the strong interaction, all unstable/resonant as well as
stable hadrons are to be included.  The propagator of a resonant
particle is given by the conventional Feynman propagator with
substitution of $ i\epsilon$ by $i\sqrt{s}\Gamma(s)$.

Four processes are considered, i) via
$K_{S}\pi^{\pm}$ resonances, $J/\psi\rightarrow
K^{*}(892)^{\mp}R_{K_{S}\pi^{\pm}},$ ii) via
$K^{*}(892)^{\mp}\pi^{\pm}$ resonances, $J/\psi\rightarrow
K_{S}R_{K^{*}(892)^{\mp}\pi^{\pm}}$, iii) via $K^{*}(892)^{\mp}K_{S}$
resonances, $J/\psi\rightarrow\pi^{\pm}R_{K^{*}(892)^{\mp}K_{S}},$ and
iv) via a direct $K^{*}(892)^{\mp}K_{S}\pi^{\pm}$ decay, where
$R_{K_{S}\pi^{\pm}}$, $R_{K^{*}(892)^{\mp}\pi^{\pm}}$ and
$R_{K^{*}(892)^{\mp}K_{S}}$ stand for the intermediate resonant states
decaying into $ K_{S}\pi$ ($\kappa$, $K_{0}^{*}(1430)$,
$K_{2}^{*}(1430)$, $K_{2}^{*}(1980)$), $K^{*}(892)^{\mp}\pi$
($K_{1}(1270)$, $K_{1}(1400)$), and $K^{*}(892)^{\mp}K_{S}$
($b_{1}(1235)$), respectively.

For the scalar $ K_{S}\pi$ resonant states$,\ \kappa$ and
$K_{0}^{*}(1430)$ are considered for $R_{K_{S}\pi^{\pm}}$. The
Lagrangian of strong interaction, $\mathcal{L}_{S}$ describing the
process is taken to be the most simple form. The Lagrangian,
$\mathcal{L}_{S}$ and corresponding amplitude, $\mathcal{F}_{S}$ are
given as follows,
\vspace{-1em}
\begin{eqnarray}
 & {\mathcal L}_S &   =  \sum_{R=\kappa ,K_0^*} ( \xi_R \psi_\mu K^*_\mu R + g_R R K_S \pi ), \nonumber \\
 & {\mathcal F}_S & =   S_{h_\psi h_{K^*}} \sum_{R=\kappa ,K_0^*} 
                          r_R e^{i\theta_R} \Delta_R (s_{K_S \pi}), \nonumber \\
 & \Delta_R&(s_{K_S \pi})  = 
     \frac{m_R\Gamma_R}{m_R^2-s_{K_S \pi}-i\sqrt s_{K_S \pi}\Gamma_R(s_{K_S \pi})},
  \label{eq2}
\end{eqnarray}
where $\Delta_R(s_{K_S \pi})$ is the Breit-Wigner formula with 
$\Gamma_R(s_{K_S \pi})= pg_R^2/(8\pi s_{K_S \pi})$, describing the 
decay of $R=\kappa$ and $K_0^*(1430)$, and $S_{h_\psi h_{K^*}} \equiv 
\epsilon_\mu^{(h_\psi )} \tilde\epsilon_\mu^{(h_{K^*})}$ is a factor 
due to helicity combinations between relevant particles. 
$p$ is a momentum of the pion decaying from the $K_{S}\pi$ system at rest. 
$S_{h_{\psi}h_{K^{*}}}r_{R}e^{i\theta_{R}}$ describes the S-matrix
element, $_{\ out}\langle RK^{*}|J/\psi\rangle_{in}$, where
$e^{i\theta_{R}}$ parameterizes the rescattering phase of the outgoing
wave$,_{\ out}\langle RK^{*}|$.  This form of $\mathcal{F}_{S}$ is
consistent with the generalized unitarity of the S-matrix.

\begin{figure}[thbp]
\includegraphics[width=14cm,angle=0]{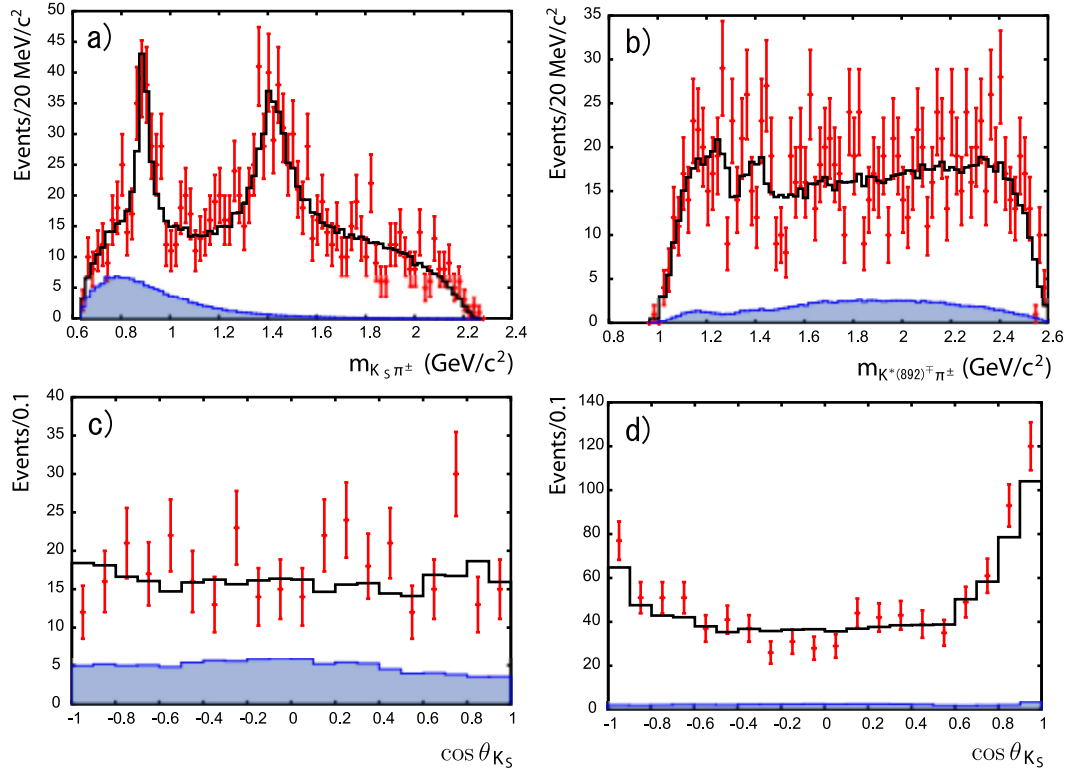}
\caption{
   {\footnotesize 
     Results of the analysis. The mass distributions of (a)
     $K_{S}\pi^{\pm}$ and (b) $K^{*}(892)^{\mp}\pi^{\pm}$ and the
     angular distributions of $K_{S}$ in the $ K_{S}\pi$ center of
     mass system for the $ K_{S}\pi$ mass regions (c) below $1 \ 
     \mbox{GeV}/c^{2}$ and (d) above $1 \ \mbox{GeV}/c^{2}$.  Crosses
     with error bars are the experimental data and solid histograms
     are fitted results. The shaded histograms (blue color) show the
     contributions from the charged $\kappa$.  }  }
\label{fig:8}
\end{figure}
The decay amplitudes through $R_{(K_{S}\pi)_D}$,
$R_{K^{*}(892)^{\mp}\pi}$ and $R_{K^{*}(892)^{\mp}K}$ denoted as
$\mathcal{F}_{D},\ \mathcal{F}_{K_{1}},$ and $\mathcal{F}_{b_{1}}$,
respectively, are obtained in a similar manner.  The direct $
K^{*}(892)K_{S}\pi$ production amplitude is taken to be
$\mathcal{F}_{direct}=S_{h_{\psi}h_{K^{*}}}r_{K_{S}\pi}e^{i\theta_{K_{S}\pi}}$.
The total amplitude, $\mathcal{F}$ is given by the sum of all
amplitudes, $\mathcal{F}=\ 
\mathcal{F}_{S}+\mathcal{F}_{D}+\mathcal{F}_{K_{1}}+\mathcal{F}_{b_{1}}+\mathcal{F}_{direct}$.
Details of the amplitudes considered in the analysis are described in \cite{had03}. 

Background events of $K^*(892)_{BG}$ decaying into $K_{S}\pi$ are
taken into the analysis.  They are events of $K^{*}(892)$ recoiling
against $ K_{S}\pi$ systems.  The $ K_{S}\pi$ system is in the mass
region of $K^*(892)$ in an alternate combination channel of the
$K^*(892)K_{S}\pi$, and is seen at the cross region of the $K^*(892)$
bands of the $K_{S}\pi$ scatter plots in Fig. \ref{fig:2}a) or
\ref{fig:2}b).  The process $K^*(892)K^*(892)_{BG}$ is described by an
amplitude being incoherent with $\mathcal{F}$.
The non-interfering phase space-like background is also considered. Its amount estimated from the $K^{*}(892)$ side-band events is fixed in the analysis. 
%


The PWA is performed on the $ K_{S}\pi$ and the $ K^{*}(892)\pi$ mass
distributions and the angular distributions of $K_{S}$ in the $
K_{S}\pi$ center of mass system. 
The BES detector simulation code, SIMBES \cite{simbes} 
which is a GEANT3-based Monte Carlo program is used for the acceptance correction. 
$\chi^{2}$ fitting is utilized in
the partial wave analysis.  
The MINUIT package of CERNLIB \cite{MINUIT} is used for minimization 
of functions and estimation of uncertainties in the fitting.
Fig. \ref{fig:8} shows the results of fitting on 
$m_{K_{S}\pi^{\pm}}$, $m_{K^{*}(892)^{\mp}\pi^{\pm}}$, 
and the angular distributions of $K_{S}$ in the $ K_{S}\pi$ center of mass system 
for the $ K_{S}\pi$ mass regions below and above $1$ GeV$/c^{2}$. 
The $ K_{S}\pi$ mass distribution and the $K_{S}$ angular
distributions are well reproduced by the fit. 
The shaded histograms (blue color) show the
contributions of the $\kappa$ resonance.

The parameter values of mass and width of the resonances except the
$\kappa$ resonance are fixed in the PWA to those summarized in the PDG
tables \cite{1}. The uncertainties of $1 \sigma$ deviations of the
resonance parameters are included in the estimation of the systematic
errors of the $\kappa$ parameters.

\begin{figure}[thbp]
\begin{center}
\includegraphics[width=12cm,angle=0]{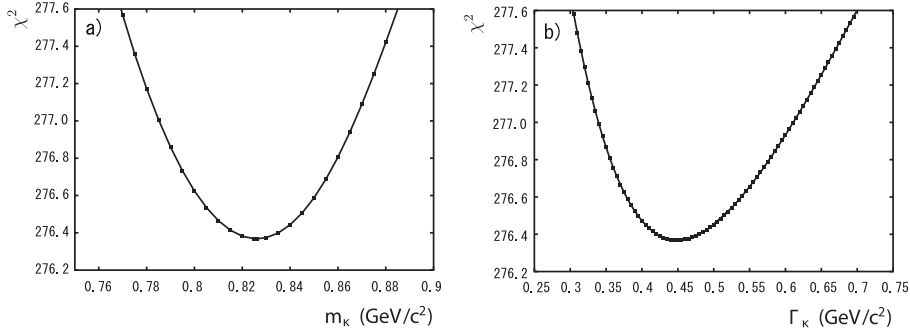}
\end{center}
\caption{
   {\footnotesize 
Mass and width scan for the charged $\kappa$, a) mass scan and b) width scan.
   }
}
\label{fig:9}
\end{figure}

Events around 1.9 GeV$/c^{2}$ in the $M_{K_{S}\pi}$ distribution
require a resonance in the fitting. Resonances, $K_{0}^{*}(1950)$ and
$K_{2}^{*}(1980)$, are tried. The significance obtained for the
$K_{2}^{*}(1980)$ is better than that for $K_{0}^{*}(1950)$ by $ 1.43 \sigma$, 
and $K_{2}^{*}(1980)$ is included in the PWA.  
The $K^{*}(892)^{\mp}K_{S}\pi^{\pm}$ direct decay which interferes with
$\kappa$ and other resonances is found to have less than $1 \sigma$
significance and is not included in the present analysis.


Uncertainties of the contribution of the non-interfering phase
space-like background are included in the estimation of the systematic
errors of the $\kappa$ parameters, assuming this background to be
uncertain by $\pm 30\%$.


Breit-Wigner parameters for mass and width of the charged $\kappa$ are
obtained to be
$$m_{\kappa}= 826 \pm 49_{-34}^{+49} \ \mbox{MeV}/c^{2} \ \ \mbox{and} \ \ 
\Gamma_{\kappa}= 449 \pm 156_{-81}^{+144} \ \mbox{MeV}/c^{2},$$ 
where the first errors are statistical and the second ones are
systematic.
The pole position of the charged $\kappa$ is determined to be
$$(764\pm 63_{-54}^{+71})-i(306\pm 149_{-85}^{+143}) \  \mbox{MeV}/c^{2}$$
from the Breit-Wigner parameters.

The $\kappa$ resonance parameters are scanned and are confirmed to
have no local minimum as shown in Fig. \ref{fig:9}. The $\chi^{2\ 
}$value of the PWA including charged $\kappa$ is 276 with 185 degrees
of freedom (d.o.f). That without the charged $\kappa$ is 344 with 189
d.o.f.  This indicates the significance of the charged $\kappa$ to
be $7.5\sigma.$

The value for each distribution, $\chi^{2}/N_{data}$ ($N_{data}$: number of data points) in the fit including the charged $\kappa$  may help to give ideas on the fitting affairs of the PWA. They are 0.99 for $\chi_{m_{K_{S}\pi}}^{2}/N_{data}$, 1.82 for $\chi_{m_{K^{*}\pi}}^{2}/N_{data}$, 1.39 for $\chi_{\cos\theta_{K_{S}}}^{2} (m_{K_{S}\pi}\leq 1\ \mathrm{G}\mathrm{e}\mathrm{V}/c^{2})/N_{data}$ and 1.09 for $\chi_{\cos\theta_{K_{S}}}^{2}(m_{K_{S}\pi}>1\ \mathrm{G}\mathrm{e}\mathrm{V}/c^{2})/N_{data}$. These values show that the contribution to the total $\chi^{2}$ comes mostly from the fit of the $ K^{*}\pi$ mass distribution.

\begin{table}[hbpt]
\caption { 
Observed values for parameters of the charged kappa and for phases of resonances. 
Mass and width values of resonances except $\kappa$ are fixed to those in the PDG tables \cite{1}. 
}
\label{tab:1}

 \begin{center}
    \begin{tabular}{lcccc}  \hline 
    
  Process  & Mass (MeV$/c^{2}$) &  Width (MeV$/c^{2}$) & $\theta$ (Deg.)  \\ \hline 
    
  $ \kappa  \rightarrow K \pi $   & 
     
$826\pm 49_{-34}^{+49}$ & $449\pm 156_{-81}^{+144}$ &    0  (fixed) \\ 
    
  $ K_0^*(1430) \rightarrow K \pi        $  & 1425 & 270 &  37 \\  
  $ K_2^*(1430) \rightarrow K \pi        $  & 1426 &  99 & 157 \\ 
  $ K_2^*(1980) \rightarrow K \pi        $  & 1973 & 373 &  16 \\ 
  $ K_1(1270)   \rightarrow K^*(892) \pi $  & 1272 &  90 & 111 \\ 
  $ K_1(1400)   \rightarrow K^*(892) \pi $  & 1403 & 174 & 111 \\ 
  $ b_1(1235)   \rightarrow K^*(892) K   $  & 1230 & 142 & 235   \vspace{1em} \\

  $ K^*(892)_{BG}  \rightarrow K \pi $  &  892 &  51 &  ---  \\ 
   Non-interfering phase space-like B.G.      &  ---   &  ---  &  ---  \\ \hline 
    
 \end{tabular}
 \end{center}
 \end{table}
\begin{table}[hbpt]
\caption { 
Numerical values of the relative contributions of the resonances and backgrounds.
They are normalized by the total contribution.
Uncertainties are statistical. 
}
\label{tab:2}
 \begin{center}
\begin{tabular}{ll} \hline
$ J/\psi \to   \kappa K^*(892)      $    & $ 0.109 ^{+0.023}_{-0.021} $ \\ 
$ J/\psi \to   K_0^*(1430) K^*(892) $    & $ 0.115 ^{+0.052}_{-0.042} $ \\ 
$ J/\psi \to   K_2^*(1430) K^*(892) $    & $ 0.114 ^{+0.038}_{-0.033}  $ \\
$ J/\psi \to   K_2^*(1980) K^*(892) $    & $ 0.017 ^{+0.018}_{-0.011}  $ \\ 
$ J/\psi \to   K_1(1270) K          $    & $ 0.044 ^{+0.015}_{-0.013} $ \\ 
$ J/\psi \to   K_1(1400) K          $    & $ 0.069 ^{+0.028}_{-0.023}  $ \\ 
$ J/\psi \to   b_1(1235) \pi        $    & $ 0.069 ^{+0.030}_{-0.024}  $ \vspace{1em} \\ 
$ K^*(892)_{BG}     $                & $ 0.084  ^{+0.015}_{-0.014} $ \\ 
Non-interfering phase space-like B.G.      & $ 0.328 $ {\scriptsize $\pm 0.098$} \\ \hline
\end{tabular}
\end{center}
\end{table}

The parameter values obtained for the charged $\kappa$ are summarized
in Table \ref{tab:1}. The mass and width values for other resonances
used in the PWA are also shown. The relative
phases obtained for resonances are in the last column. The parameter
values obtained for the charged $\kappa$ are in good agreement with
those for the neutral $\kappa$ \cite{12}, 
$m_{\kappa}=(878\pm23_{-55}^{+64}) \ \mbox{MeV}/c^{2}$ and 
$\Gamma_{\kappa}=(499\pm 52_{-87}^{+55}) \ \mbox{MeV}/c^{2}$.

Table \ref{tab:2} shows the relative contributions of the relevant resonances and of the backgrounds from $K^{*}(892)_{BG}$ and the non-interfering phase space-like background. 
The relative contribution is obtained by integration of the amplitude squared 
of the relevant resonance or background with normalization by the total contribution which is from a sum of the total amplitude squared and the background amplitudes squared.

A constant width parameterization of the Breit-Wigner formula
for each resonance is also tried in the PWA. The mass and
width values of $\kappa$ are obtained to be $m_{\kappa}=656\pm
119_{-186}^{+56}$ MeV/$c^{2}$ and $\Gamma_{\kappa}=536\pm
211_{-144}^{+440}$ MeV$/c^{2}$, and the parameters for the pole position
are determined to be $(702\pm 113_{-175}^{+54})-i(250\pm
88_{-62}^{+163})$ MeV$/c^{2}$. They are consistent with those obtained
for the s-dependent width in Eq. (\ref{eq2}).

Recently the BES Collaboration reported \cite{22} the mass and width for a
charged $\kappa$ in $K_{S}\pi^{\pm}$ and $K^{\pm}\pi^{0}$ recoiling against
$K^{*}(892)^{\mp}$ in $J/\psi\rightarrow K^{\pm}K_{S}\pi^{\mp}\pi^{0}$ 
to be ($884\pm 40_{-22}^{+11}$) MeV$/c^{2}$ and ($478\pm 77_{-41}^{+71}$) MeV$/c^{2}$, 
respectively.
The parameter values of the $\kappa$
resonance are in good agreement with those of the neutral $\kappa$
\cite{12} and also with those of the present results.

A possible contribution from $K^{*}(892)^{\pm}$ recoiling against
$K^{*}(892)^{\mp}$ is studied.  This process may proceed via one
photon annihilation of the $ J/\psi$ decay.  Including both
$K^{*}(892)^{\pm}$ and the background, $K^{*}(892)_{BG}^{\pm}$, the
PWA is unable to determine their amplitudes separately due to the low
statistics of the present data.  Also the PWA with
$K^{*}(892)^{\pm}$ instead of $K^{*}(892)_{BG}^{\pm}$ finds 
almost the same results for the mass, width, amplitude, and 
phase values of the 
$\kappa,\ K_{0}^{*}(1430)$, $K_{2}^{*}(1430)$ and $K_{2}^{*}(1980)$, and 
slightly different results of $K_{1}(1270),\ K_{1}(1400)$ and $b_{1}(1235)$. 
The resonance state, $K^{*}(1410)^{\pm}$ is also examined in the PWA.  The results
for $K^{*}(1410)^{\pm}$ are consistent with
zero with little effect on the parameters of other
resonances. 
The process, $K^{*}(1410)$ recoiling against $K^{*}(892)$ in the
$J/\psi$ decay is definitely suppressed.

The branching ratio for $J/\psi\rightarrow
K^{*}(892)^{\mp}\kappa^{\pm}$ is determined.
The number of events for $K^{*}(892)^{\mp}\kappa^{\pm}$ is
determined to be $142\pm 28$ in the PWA, and the number of $J/\psi$
events collected by BESII is $(5.8\pm 0.3)\times 10^{7}$.  The
detection efficiency is estimated to be $4.88\times 10^{-3}$ by Monte
Carlo simulation.  It includes the detector efficiencies and isospin
factors of the decays, $K^{0}\rightarrow K_{S}$ and
$K_{S}\rightarrow\pi^{+}\pi^{-}$.  The branching ratio for the
process, $BR(J/\psi\rightarrow K^{*}(892)^{\mp}\kappa^{\pm}$) is
determined to be
\begin{eqnarray}
  BR(J/\psi\rightarrow K^{*}(892)^{\mp}\kappa^{\pm}) 
    & = &
      \frac{142}{(5.8\times 10^{7})(4.88\times 10^{-3})}\times\frac{9}{4} \nonumber \\ 
    & = &
      (1.13\pm 0.22_{-0.22}^{+0.49})\times 10^{-3}, \nonumber
\end{eqnarray}
where the first error is statistical and the second systematic.  The
factor, $\frac{9}{4}$ comes from the isospin weight for the decay
modes of $\kappa$ and $K^{*}(892)$.  The systematic errors are
estimated by square root of the sum of uncertainties for the number of
$K^{*}(892)^{\mp}\kappa^{\pm}$ events obtained in PWA, for the
detection efficiency in Monte Carlo simulation, and for the number of
$J/\psi$ events.  
The branching ratio is also reported by the BES
Collaboration \cite{22} in the analysis of $J/\psi\rightarrow
K^{*}(892)^{\mp}\kappa^{\pm}\rightarrow K^{\mp}K_{S}\pi^{\pm}\pi^{0}$ 
to be $(1.09\pm 0.18^{+0.94}_{-0.54})\times10^{-3}$, which 
is consistent with the present result.


In summary the charged $\kappa$ is observed in the PWA analysis of the
$K_{S}\pi^{\pm}$ system recoiling against $K^{*}(892)^{\mp}$ in the
decay, $J/\psi\rightarrow K^{*}(892)^{\mp}K_{S}\pi^{\pm}\rightarrow
K_{S}K_{S}\pi^{+}\pi^{-}$.  The $K_{S}K_{S}\pi^{+}\pi^{-}$ data are
selected from six charged track events of the 58 million $ J/\psi$
decays obtained by BESII at BEPC. 
Contributions from backgrounds in the selections of 
the $K_{S}K_{S}\pi^{+}\pi^{-}$ events and of 
the $K^{*}(892)^{\mp}K_{S}\pi^{\pm}$ events are considered.
The partial wave analysis is
performed based on the VMW method, and $\chi^{2}$ fitting is utilized
for the fit. Breit-Wigner parameters for the charged $\kappa$
resonance are obtained in the analysis.  The mass and width parameters
of the resonances contributing the process except the charged $\kappa$
are fixed in the analysis to those of the PDG tables.  Their
uncertainties and those from the estimation of the backgrounds are
included in the estimation of the systematic errors for the charged
$\kappa$ parameters.  The parameters of the charged $\kappa$ are in
agreement with those for the neutral $\kappa$.  The branching ratio of
charged $\kappa$ of the decay, $J/\psi\rightarrow
K^{*}(892)^{\mp}\kappa^{\pm}$ is obtained.  The result is consistent
with that obtained recently at BESII for the charged $\kappa$ in the
different channels of the $J/\psi$ decays.


{\bf Acknowledgments}

The BES Collaboration thanks the staff of BEPC and computing
center for their hard efforts. This work is supported in part by
the National Natural Science Foundation of China under contracts
Nos. 10491300, 10225524, 10225525, 10425523, 10625524, 10521003, 10821063,
10825524, the Chinese Academy of Sciences under contract No. KJ 95T-03, the
100 Talents Program of CAS under Contract Nos. U-11, U-24, U-25,
and the Knowledge Innovation Project of CAS under Contract Nos.
U-602, U-34 (IHEP), the National Natural Science Foundation of
China under Contract Nos. 10775077, 10225522 (Tsinghua University), 
the Core University Program of Japan Society for Promotion of Science, 
JSPS under contract No. JR-02-B4 (KEK and Universities), the fund for 
the international collaboration and exchange (Nihon-U), Grants-in-Aid 
for Science Research under contract No. C18540281 (U. Miyazaki), 
and the Department of Energy under Contract No. DE-FG02-04ER41291 (U. Hawaii).



\end{document}